\documentclass[preprint,showkeys,aps,prb]{revtex4}
\usepackage[dvips]{graphicx}
\begin{document}

\title{The Nos\'e-Hoover, Dettmann, and Hoover-Holian Oscillators}

\author{
William Graham Hoover,                 
Ruby Valley Research Institute\\        
601 Highway Contract 60,
Ruby Valley, Nevada 89833;\\
Julien Clinton Sprott,                 
Department of Physics\\                 
University of Wisconsin-Madison,       
 Madison, Wisconsin 53706;\\     
Carol Griswold Hoover,                  
Ruby Valley Research Institute\\        
601 Highway Contract 60,               
Ruby Valley, Nevada 89833.         
}

\date{\today}

\vspace{0.1cm}

\begin{abstract}
To follow up recent work of Xiao-Song Yang\cite{b1} on the Nos\'e-Hoover oscillator\cite{b2,b3,b4,b5}
we consider Dettmann's harmonic oscillator\cite{b6,b7}, which relates Yang's ideas directly to
Hamiltonian mechanics. We also use the Hoover-Holian oscillator\cite{b8} to relate our mechanical studies
to Gibbs' statistical mechanics. All three oscillators are described by a coordinate $q$ and a momentum $p$.
Additional control variables $(\zeta, \xi)$ vary the energy. Dettmann's description includes a time-scaling
variable $s$, as does Nos\'e's original work\cite{b2,b3}.  Time scaling controls the rates at which the
$(q,p,\zeta)$ variables change. The ergodic Hoover-Holian oscillator provides the stationary Gibbsian
probability density for the time-scaling variable $s$. Yang considered {\it qualitative} features of
Nos\'e-Hoover dynamics.  He showed that longtime Nos\'e-Hoover trajectories change energy, repeatedly
crossing the $\zeta = 0$ plane. We use moments of the motion equations to give two new, different, and brief
proofs of Yang's long-time limiting result.

\end{abstract}

\maketitle

\section{Background}

Nos\'e-Hoover dynamics was developed in 1984 as a side benefit of a Centre Europ\'een de Calcul
Atomique et Mol\'eculaire (``CECAM'') Workshop on Constrained Dynamics organized by Carl Moser.
The workshop was held at Orsay, about 20 miles southwest of Paris.  Shuichi Nos\'e and Bill
Hoover met by chance at the Orly airport a few days prior to the Orsay meeting and were able to
spend several hours together near the Notre Dame cathedral, discussing Nos\'e's recent work on the
thermal control of molecular dynamics simulations\cite{b2,b3}.

Seeking better to understand Nos\'e's innovative work Hoover applied Nos\'e's ideas to the
one-dimensional harmonic oscillator problem\cite{b4,b5,b6,b7,b8,b9,b10,b11,b12}. He generated
hundreds of solutions of the set of three ordinary differential equations using the fourth-order
Runge-Kutta algorithm. The time-dependent variables $(q,p,\zeta)$ are respectively the oscillator
coordinate, momentum, and friction coefficient. Here are the three equations :
$$                                
\{ \ \dot q = p \ ; \ \dot p = -q -\zeta p \ ; \ \dot \zeta = [ \ p^2 - 1 \ ]/\tau^2 \ \} \
[ \ {\rm Nos\acute{e}-Hoover} \ ] .
$$
Provided that the mean value of the friction coefficient $\zeta$ is finite ( so that the longtime
average  of its time derivative, $\langle \ \dot \zeta \ \rangle$, vanishes ) the last equation
implies that the mean oscillator temperature $\langle \ p^2 \ \rangle$ is unity for long times.
Thus $\zeta$ acts as a ``thermostat''. Posch, Hoover, and Vesely explored the detailed nature of
the oscillator solutions\cite{b5} for a variety of response times $\tau$.  Wang and Yang extended
this work\cite{b9,b10} and found, in addition to the known periodic, toroidal, and fat-fractal
chaotic trajectories, additional periodic trajectories, and tori, in the form of knots ! The types
of solutions found vary with $\tau$ and with the initial values of $(q,p,\zeta)$.  In the present
work we choose $\tau = 1$.

\section{Consequences of Time Scaling}

In the Summer of 1996 Carl Dettmann discovered a {\it vanishing} Hamiltonian which precisely
reproduces the time-dependence of the Nos\'e-Hoover flow where the Nos\'e-Hoover momentum becomes
$(p/s)$\cite{b6,b7}.
$$
{\cal H}_D(q,p,s,\zeta) \equiv (s/2)[ \ q^2 + (p/s)^2 + \ln(s^2) + \zeta^2  \ ] \equiv 0 \rightarrow
$$
$$
s^2 = e^{-(q^2 + (p/s)^2 + \zeta^2)} \ .
$$
The ``time-scaling variable'' $s$ has the conjugate momentum $p_s = \zeta$. $| s |$ must be less than
unity for ${\cal H}_D$ to vanish. We select positive values, $0 < s < 1$, in our numerical work. The
various resulting flows can occupy one-, two-, or three-dimensional subspaces of the full stationary
solution, the Gaussian function :
$$
(2\pi)^{-3/2}\exp [ \ -( q^2 + (p/s)^2 + \zeta^2 )/2 \ ] \ .
$$
With this Gaussian distribution known the normalized probability density for $s$ follows easily,
$$
\sqrt{(2/\pi)\ln(s^{-2})} \ .
$$

\begin{figure}
\includegraphics[width=3.0in,angle=+90.]{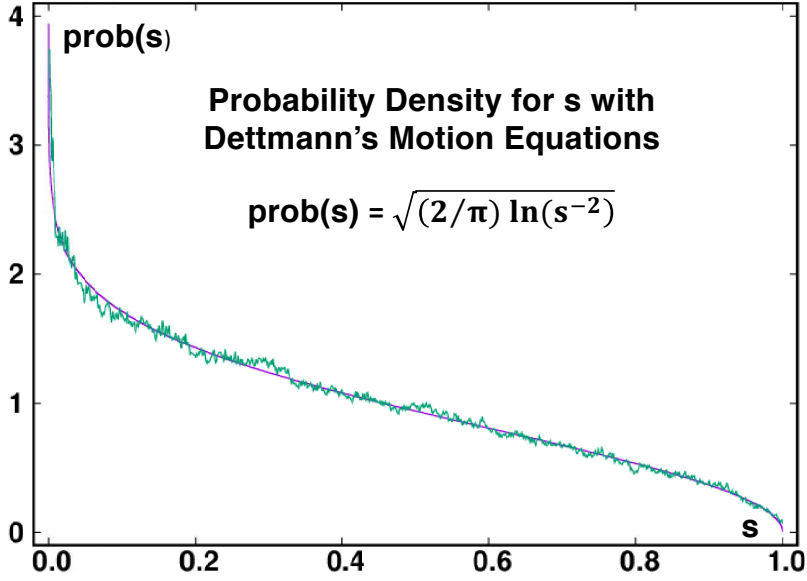}
\caption{
Analytic and numerical probability densities for the time-scaling variable $s$. Both are based on
the three-dimensional Gaussian distribution with numerical values from the Hoover-Holian model.
}
\end{figure}
Figure 1 compares this analytic distribution to a thousand-point histogram from a fourth-order
Runge-Kutta solution of the ( ergodic ) Hoover-Holian oscillator equations\cite{b8} :
$$
\{ \ \dot q = p \ ; \ \dot p = - q - \zeta p - \xi p^3 \ ; \ \dot \zeta = p^2 - 1
\ ; \ \dot \xi = p^4 - 3p^2 \ \} \ .
$$
For the Figure  $q$ and $p$ were initially unity and the two control variables were zero.
Ten million timesteps, $dt = 0.001$, were used.  The stationary distribution for the Hoover-Holian
equations is the four-dimensional Gaussian $(2\pi)^{-2}\exp[ \ ( -q^2 - p^2 -\zeta^2 -\xi^2 )/2 \ ] \ $.
Because the model is ergodic, covering the entire four-dimensional Gaussian\cite{b8}, we were able to use
the three-dimensional Hoover-Holian subset $( \ q^2,p^2,\zeta^2 \ )$ data as a model for the stationary
Dettmann distribution $( \ q^2,(p/s)^2,\zeta^2 \ )$ with $s$ determined from the condition ${\cal H}_D 
\equiv 0$. These numerical data, confirming our analytic work, are shown in Figure 1. 
 
Following up his earlier investigations of oscillator trajectories ( including knots\cite{b9,b10}
! ) Professor Xiao-Song Yang has recently proved that the oscillator motion, independent of the
chosen initial conditions, must nearly always cross the $\zeta = 0$ plane\cite{b1}.  Yang 
considered the Nos\'e-Hoover equations, which follow from Dettmann's Hamiltonian with the
replacement $(p/s) \rightarrow p$ :
$$
\{ \ \dot q = p \ ; \ \dot p = - q - \zeta p \ ; \ \dot \zeta = p^2 - 1 \ \} \ .
$$
The only situation in which the $\zeta = 0$ plane is not crossed repeatedly is an unstable 
straight-line portion of the $\zeta$ axis :
$$
q(t) = q(0) = 0 \ ; \ p(t) = p(0) = 0 \ ; \ \zeta(t) = \zeta(0) - t \ .
$$

In the present work we first consider oscillator trajectories from the standpoint of the continuity
equation in $(q,p,\zeta)$ space. Because the Nos\'e-Hoover equations are not strictly Hamiltonian,
except in the ${\cal H}_D \equiv 0$ case discovered by Dettmann, the Nos\'e-Hoover flow is
compressible -- the three-dimensional divergence is locally nonzero, responding linearly to the
control variable $\zeta$ :
$$
(\partial \dot q/\partial q) + (\partial \dot p/\partial p) +
(\partial \dot \zeta/\partial \zeta) = 0 -\zeta + 0 \ [ \ {\rm Dettmann=Nos\acute{e}-Hoover} \ ] \ .
$$
Consider a longtime trajectory. Evidently a {\it positive} longtime average of $\zeta$ would
correspond to a {\it vanishing of the comoving volume} and a dimensionality loss. A negative average
would correspond to divergence, a numerical instability.  In fact, the continuity equation has been
used to explain the fractal nature of chaotic flows with positive friction\cite{b12}.

In $(q,p,\zeta)$ space the continuity equation shows that the comoving volume element $\otimes$
vanishes and can become fractal if the longtime-averaged control variable is positive. The volume
diverges, and the simulation stops, if that averaged variable is negative. The motion equations for
the oscillator, in either $(q,p,\zeta)$ space or $(q,p,s,p_s=\zeta)$ space, can also be used to prove
Professor Yang's zero-crossing result.  We turn to that next.

\section{Proofs of Professor Yang's $\zeta = 0$ Plane Result}

Yang's recent contribution\cite{b1} considers the three-variable Nos\'e-Hoover oscillator and
shows algebraically that any longtime trajectory -- other than the special constant $(q,p,\dot \zeta)$
unstable straight line -- must repeatedly cross the $\zeta = 0$ plane.  This is a handy result as the
Poincar\'e sections at that plane are commonly used to diagnose the fractal character of nonlinear
flows. Yang's proof-of-crossing is relatively long, thirty pages, mainly algebra. Let us provide two
simpler demonstrations of his result. Multiply the three motion equations by $q$, $p$, and $\zeta$
respectively and compute their longtime $(t \rightarrow \infty)$ averages $\langle \ \dots \ \rangle$
in the usual way :
$$
\langle \ f[q,p,\zeta] \ \rangle =
(1/t)\int_0^{t\rightarrow \infty} f[ \ q(t^\prime),p(t^\prime),\zeta(t^\prime) \ ]dt^\prime \ .
$$
Assume also that 
$\langle \ q^2,p^2,\zeta^2 \ \rangle$ are finite ( so that their time derivatives average to zero ).
The three Nos\'e-Hoover differential equations then give three identities:
$$
\{ \
  \langle \ qp \ \rangle = 0 \ ;
\ \langle \ qp \ \rangle = \langle \ -\zeta p^2 \ \rangle \ ;
\langle	  \ \zeta \ \rangle = \langle	       \ \zeta p^2 \ \rangle \ 
\ \} \ .
$$
Combining the three shows that the mean value of $\zeta$ vanishes, equivalent to Professor Yang's
result.

An even simpler demonstration follows from Nos\'e's original Hamiltonian\cite{b2,b3}, ${\cal H}_N \equiv
(1/s){\cal H}_D$, which gives a simple evolution equation for $s$ : $\dot s = p_s$. Because
$p_s$ and $\zeta$ are identical, as shown by Dettmann, the time-average of the $\dot s$ equation shows directly (assuming numerical convergence
of the motion) that the control variable $\zeta$ has mean value zero, again implying Yang's result.

\section{An Afterword for Young Researchers}

Lingering geometric and topological questions remain where the chaotic sea is concerned. The
sea is an enduring paradoxical concept. Is it a set of three-dimensional points? Is it a
continuum with holes here and there? Is it just a single chaotic trajectory? Is it an ill-defined
limit? These questions remain at the research frontier.  Let us have a look at the chaotic sea
for the Nos\'e-Hoover oscillator.

{\bf Figure 2} shows cross-sections of the three-dimensional oscillator's $(q,p,\zeta)$
phase space showing about ten million crossings of the three planes where each of the variables
vanishes.  The fourfold symmetry of these sections reflects the time-reversibility of the motion
equations as well as the fact that any solution $\{ \ +q,+p,+\zeta \  \}_t$ implies the existence of
a mirror-image solution $\{ \ -q,-p,+\zeta \  \}_t$, as well as $\{ \ +q,-p,-\zeta \ \}_t$.

\begin{figure}
\includegraphics[width=2.0in,angle=-90.]{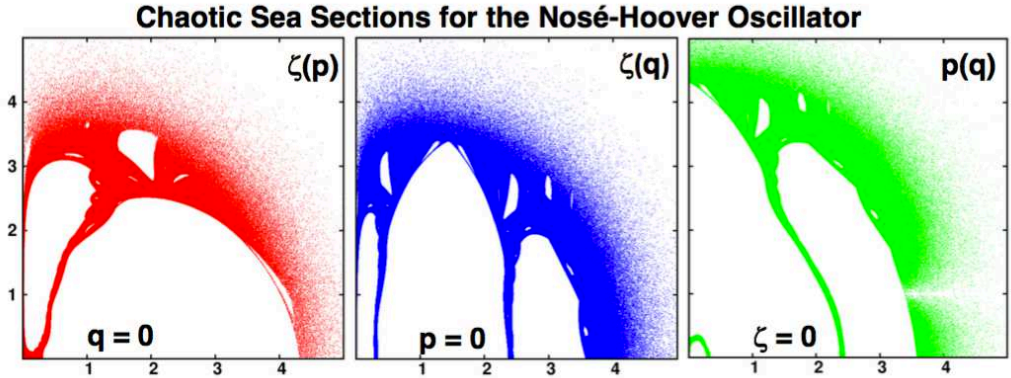}
\caption{
Sections of the Nos\'e-Hoover Chaotic Sea where $q$ or $p$ or $\zeta$ vanishes.  Only one quadrant is
shown here as each of the Sections has fourfold symmetry.
}
\end{figure}

For $\tau = 1$ about six percent of the Gaussian distribution\cite{b13} $e^{-(q^2+p^2+\zeta^2)/2}$
appears to be a cohesive connected sea, penetrated by infinitely many regular orbits with zero
Lyapunov exponents $\{ \ \lambda \ \}$. These exponents describe the longtime average growth or
decay rates of small perturbations. The boundary between points with a positive $\lambda$,
charateristic of the sea, and points with all zero Lyapunov exponents is evidently murky and
uncertain. The boundary region is likely fractal in the sense that in the neighborhood of any
chaotic point in the sea there must be other points only a small distance away, located on
periodic orbits of great length, well beyond our capacity to compute in any meaningful way.

The mathematics of such sets of points is confused through concepts related to Hilbert's Hotel and
the Banach-Tarski Theorem.  Hilbert's Hotel, with positive integer room numbers, is alway full
( with $\aleph_0$ guests ) but with always an available room ( by moving customers from n to n+1 ).
Similar mappings of points in two or three ( or many ) dimensional spaces enable the disection of
a sphere into a finite number of pieces which can then be reassembled to make two spheres. This is
an active field of mathematics at the moment. For a nicely illustrated description of the
Banach-Tarski paradox see Reference 14.

\section{Acknowledgement}

We thank Professor Xiao-Song Yang for useful email communications and the anonymous referee
for several useful suggestions which have improved this manuscript.

\end{document}